\documentclass{article}
\usepackage{graphicx}
\usepackage{url}

\begin{document}
\title{Efficient Social Choice via NLP and Sampling}
\author{Lior Ashkenazy and Nimrod Talmon \\ Ben-Gurion University}
\maketitle

\begin{abstract}
Attention-Aware Social Choice tackles the fundamental conflict faced by some agent communities between their desire to include all members in the decision making processes and the limited time and attention that are at the disposal of the community members. Here, we investigate a combination of two techniques for attention-aware social choice, namely Natural Language Processing (NLP) and Sampling. Essentially, we propose a system in which each governance proposal to change the status quo is first sent to a trained NLP model that estimates the probability that the proposal would pass if all community members directly vote on it; then, based on such an estimation, a population sample of a certain size is being selected and the proposal is decided upon by taking the sample majority. We develop several concrete algorithms following the scheme described above and evaluate them using various data, including such from several Decentralized Autonomous Organizations (DAOs).
\end{abstract}

\section{Introduction}

We consider the problem of Attention-Aware Social Choice, in which the desire of a digital community to govern itself through the participation of its community members conflicts with the limited resources of time and attention that are at the disposal of the community members.
Indeed, direct democracy has a scalability limitation as community members cannot be subjected to a huge amount of decisions, which would reduce the quality of the individual decision making and thus of the collective decisions that are being made. Thus, there is a need for a decentralized governance system that is efficient, reliable, and resilient, at scale~\cite{field2018decentralized}.
Below we list several natural approaches for attention-aware social choice.

\begin{description}

\item[Sampling-based solutions]
These are based on randomly choosing subsets of the community as ad-hoc committees. 
E.g., a simple sampling-based solution operating on a community of $n$ agents proceeds by assigning a number of agents that are chosen uniformly at random for each proposal to change the status quo, allows these agents to vote on whether to accept or reject the proposal, and decides using majority vote.

\item[Delegation-based solutions]
These are based on allowing voters to delegate certain decisions (possibly transitively as in liquid democracy~\cite{blum2016liquid}).

\item[Cryptoeconomic solutions] 
Here, economic incentives are aligned to cause certain desired behavioral properties of the system
follow from rational agents' actions (e.g., holographic consensus~\cite{faqir2021scalable}).

\item[Solutions based on machine learning (ML)]
ML can be used to learn agent and community preferences regarding certain proposals and draw conclusions about future proposals. Concretely, a trained model can suggest governance actions for community members, or, in an extreme (and possibly dangerous) setting, a system can completely replace agent participation in the community decision-making process.

\end{description}

The focus of this paper is on the evaluation of a combination of machine learning techniques and sampling-based solutions.
Our approach is essentially as follows:
we use techniques from machine learning (ML) and natural language processing (NLP) to develop a trained model that is able to take a textual governance proposal to change the status quo and estimate the probability that the community would accept the proposal if directly voted on it; then, based on the estimation of such a trained model, we select a numerical value representing the fraction of the community that we then ask to actively vote on the proposal. We then take the sampled votes and use majority voting to decide on the proposal.
At a glimpse, our solution is composed of an estimation module -- that predicts an acceptance probability for a given proposal; a sampling module -- that, based on the prediction of the estimation module, selects a vote sample; and a decision module that decides on the fate of the proposal based on the vote sample. (A more detailed description of the architecture is given in Section~\ref{section:architecture}.)

Our aim is to evaluate the above-mentioned architecture as a solution approach to attention-aware social choice.
To this end, we develop several variants of the architecture and evaluate their performance with respect to two data sources:
  data obtained from Kaggle\footnote{\url{https://www.kaggle.com/}} (a popular venue for ML-related data) as well as data from Snapshot\footnote{\url{https://snapshot.org/}} (a popular voting application for digital communities).

\section{Preliminaries}

We provide preliminaries regarding sampling-based techniques in computational social choice as well as some technical background to NLP.

% \subsection{Decentralized Governance at Scale}
% %
% The problem of AAA is anchored at the conflict between the desire for high decision-making capacity and the limited time, attention, and effort that governance participants, i.e., community members, can devote to such processes.
% %
% Our general mathematical model contains the following ingredients:
% %
% \begin{itemize}

% \item  
% A set $A = \{a_1, \ldots, a_n\}$ of agents.
    
% \item  
% A set $P = \{p^1, \ldots, p^m\}$ of proposals, each of which is a binary proposal to change the status quo that can be either \emph{accepted} or \emph{rejected}.
    
% \item  
% The opinions of the agents, denoted by $s_i^j$, $i \in [n]$, $j \in [m]$, where $s_i^j$ is equal to $1$ ($0$) if agent $a_i$ is in favor (respectively, against) proposal $p^j$.
    
% \end{itemize}

\subsection{Sampling-Based Techniques in Social Choice}

There are some works that consider sampling-based techniques for decision making processes. For example, Bhattacharyya and Dey~\cite{dey2015sample} study sampling-based algorithms for computing election winners when votes arrive in a stream. Dey et al.~\cite{dey2017proportional} consider similar algorithms but for multiwinner voting rules that aim at proportional representation. 

Our use of sampling is as one module in our overall architecture that aims at solving attention-aware social choice.

\subsection{Natural Language Processing}

Natural Language Processing (NLP) -- a sub-field of artificial intelligence and linguistics -- deals with problems that relate to the processing, manipulation, and understanding of natural language to allow for computer reasoning on text~\cite{gobinda2003natural}. Deep learning models are very popular in NLP, although they require massive amounts of labeled data; indeed, assembling this sort of big data set is one of the main challenges of NLP~\cite{blum1998combining,dahlmeier2017challenges}. 
Some of the main functions of NLP algorithms are: text classification -- involving categorization of text by tagging; text extraction -- involving text summarization; machine translation; and more. NLP employs both \emph{syntactic analysis} -- based on the arrangement of words in a sentence; and \emph{semantic analysis} -- based on analyzing the use of and the meaning behind words.

\paragraph{Words as Features}
In NLP, it is likely that some words in a large text corpus are very prevalent (e.g. ''the'', ''a'', and ''is'' in English), thus conveying very little information about the actual content of a document. By feeding the count data directly to a classifier, those very frequent terms may ''obscure'' the frequencies of rarer, yet more interesting terms. To re-weight the count features into values that are suitable for usage by a classifier it is very common to use the TF-IDF transform~\cite{ramos2003using}, a numerical statistic that reflects how important a word is to a document in the collection or corpus~\cite{salton1988term}. The TF-IDF value grows proportionally to how often a word appears in a document, but is offset by the word's frequency (the raw frequency of a term in a document) in the corpus, which helps in reducing the effect of some words being more common than others. 

\paragraph{Words as Vector Spaces}
A more in-depth approach to NLP involves capturing the meaning of words in vector spaces, along with modeling their compositionality, hierarchy, and recursion. Given a vocabulary of words $W$, a classical NLP approach is to define a $|W|$-dimensional vector with all entries set to $0$ except one entry that identifies a word $w_t \in W$ (a fixed vocabulary of words); this is called \emph{one-hot encoding}~\cite{lebret2016word}. The disadvantage of such a representation is that it does not consider the semantic features of words, and it is rather voluminous and redundant as each word is represented by a vector~\cite{koroteev2021bert}.
A related approach is the \emph{bag of words} model. Here, the entire text is represented by a vector of size $|W|$, where each component represents the number of times that each word occurs in the text. Note that also this model does not take into account word semantics~\cite{koroteev2021bert}.
A more involved solution consists of the encoding of words into dense vectors that capture the syntactic and semantic properties of words, allowing related words to be close in a corresponding metric space. Such representation, in a space whose dimension is low, compared to the size $|W|$ of the vocabulary, is called a \emph{word embedding}~\cite{lebret2016word,alsentzer2019publicly}.

\paragraph{Language Models}
More advanced text models use \emph{transformers} to represent text~\cite{lin2021survey}: these are neural networks that allow the representation of the token (word) to be influenced by the context~\cite{bracsoveanu2020visualizing}. 
In particular, \emph{Bidirectional Encoder Representations from Transformers} (BERT) is a transformer-based ML technique for NLP~\cite{koroteev2021bert}. It provides powerful solutions for contextualized word representations and can create a context-sensitive embedding for each word in a sentence. 
% BERT is deeper and contains much more parameters than other similar models, thus possessing greater representation power. Another advantage of BERT is that, rather than simply providing word embeddings as features, it can be incorporated into a downstream task and gets fine-tuned as an integrated task-specific architecture~\cite{alsentzer2019publicly}.
%
Contextual word embedding models such as BERT have dramatically improved performance for many NLP tasks~\cite{reimers2019sentence}. 
%The BERT model training process includes two stages: pre-training on unlabeled data, and additional training on labeled data for a specific application problem. As a result, BERT can be used for text classification tasks. It is allows additional training of the basic model for a specific task using just one additional layer of neurons. This model makes it possible to obtain models with currently best performance in text classification problems~\cite{koroteev2021bert}.

\section{The Overall Architecture}\label{section:architecture}

\begin{figure}[t]
    \centering
    \includegraphics[width=11cm]{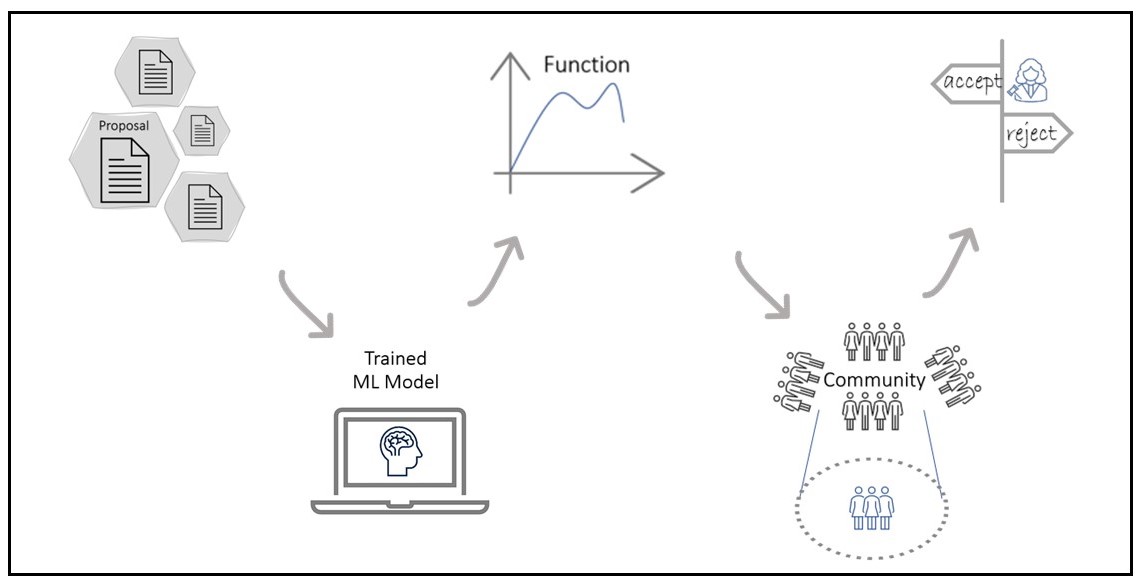}
    \caption{The Overall Architecture: when a proposal to change the status quo arrives, it goes to the trained ML model that outputs its estimation on the probability that the proposal would have passed if voted on by the community at large. Based on this prediction, a predefined function selects a fraction of the community to then actively vote on the proposal. The fate of the proposal is decided based on the majority of the sampled votes.}
    \label{figure:the overall architecture}
\end{figure}

We discuss our overall architecture, which consists of an estimation module, a sampling module, and a decision module.
A graphical representation of our solution architecture is given in Figure~\ref{figure:the overall architecture}.

% As we already know, There are many proposals that can reach the community; those proposals will be the input of the first part of the overall system which we called the estimation module. 

\paragraph{The prediction module}
The prediction module gets as input a proposal and outputs an estimation regarding whether the proposal would have passed if all community members directly vote on it.
This estimated probability is used as the input for the sampling module.

In our realizations of the architecture, the predition module is implemented as an ML model that is pre-trained on a corpus of textual proposals that are labeled by whether the proposal was accepted or rejected.

\paragraph{The sampling module}
The sampling module gets the estimated probability of the estimation module as input and, based on it, selects a subset of the votes to the directly vote on that proposal.
The essence and purpose of the sampling module is to reduce the number of community members that actively need to vote on a given proposal, thereby limiting the required community attention. 

In our realizations of the general architecture, we use different functions that take as input the probability that the estimation module outputs, and return a fraction of the population that shall be sampled; then, such a population fraction is sampled uniformly at random. Intuitively, the larger the sample size, the more accurate and quality the decision will be. However, the larger the sample size, the higher the community attention used. As our goal is to reduce the attention and effort needed for the decision making process, we explicitly consider the trade-off between the usage of community attention (i.e., the average sample size) and the quality of the decision-making process.

\paragraph{The decision module}
The decision module collects the votes of the vote sample that is selected by the sampling module and decides o the fate of the proposal. 

In our realizations we decide according to the vote majority of the sample.

\section{The Estimation Module} 

Next we discuss our specific realizations of the estimation module. 
Generally speaking, we use ML models that we train on a corpus of labeled proposals, based on NLP techniques.
%
% Recall that the estimation module is used to predict the probability that a particularly unseen proposal will be accepted if all the community members would actively vote on it. So how does it actually work? This tool is assumed to be used by digital community once the community has been operational and based for a while and has enough proposals to learn from. After applying the assumption, our model is trained on prior data, studies it, and predicts whether an existing proposal will pass and at what probability. 
%
First, given the raw textual data, we have chosen two approaches, one simple approach that is based on TF-IDF, and another that is based on BERT.

\subsection{SimpleNLP: An Initial NLP Algorithm}

As a starting point, we implemented a simple ML model that converts raw data into a matrix of TF-IDF features, indicating how important a word is to a corpus, and incorporates them into a classification algorithm. Moreover, in order to improve those models, we performed a tuning process for some of the hyper-parameters of each model using grid search.
%\footnote{Grid-search is used to find the optimal hyper-parameters of a model which results in the most ‘accurate’ predictions. The parameters of the estimator used to apply these methods are optimized by cross-validated grid-search over a parameter grid.} 
%
The following four classic algorithms have been chosen for the classification task:
\begin{description}

\item 
\textbf{Random Forest Classifier}:
This classifier consists of a combination of tree classifiers where each classifier is generated using a random vector sampled independently from the input vector, and each tree casts a unit vote for the most popular class to classify an input vector~\cite{pal2005random}.
The tuning was made on the hyper-parameters: \textit{n estimators}, which is the number of trees in the forest; and \textit{max depth} i.e. the maximum depth of the tree. 

\item
\textbf{Multinomial Naive Bayes Classifier}: 
This classifier belongs to a family of simple probability analyzers based on the Bayes theorem and is based o the assumption that the value of specific features is always different from the value of other features. There are several notable advantages to this classifier -- efficiency in time, memory, and CPU usage, as well as higher quality for small training sets~\cite{setyawan2018comparison}.
Technically, this is a generative model: 
  it assumes that each document (i.e., proposal) is generated by selecting some class for it and then generating each word of that document independently according to a class-specific distribution~\cite{xu2017bayesian}.
  
The tuning of this classifier was made on the hyper-parameters: \textit{alpha}, which is the additive smoothing parameter; and \textit{fit prior}, which decides whether to learn prior probabilities or not. 

\item
\textbf{Logistic Regression}:
This model is a general linear regression model. Its technique allows for the examination of the impact of numerical factors on binary responses. The logistical function, also known as the sigmoid function, is used to calculate the logistic model in which the output is in the range between $0$ and $1$. Due to its capability to understand vector variables and evaluate the coefficients or weights of each input variable, this model can be used in ML applications~\cite{setyawan2018comparison}.

Its tuning was made on the hyper-parameters: \textit{C}, which is the inverse of regularization strength (smaller values specify stronger regularization); and \textit{class weight}, which means the weights associated with classes. 

\end{description}

\subsection{AdvancedNLP: An Improved NLP Algorithm}

For improving SimpleNLP, we wanted to represent the data (in particular, the words) in a better way; specifically, in a way that takes into account both the meaning of the words as well as their context. This was done using BERT, which provides strong solutions for representing contextualized words.   
Next we explain the steps we worked on when implementing the BERT model. 

First, we processed our text by removing special characters and entity mentions; then we used the BERT tokenizer, which prepares the inputs for the model and as part of that has a particular way of handling words that are not in the vocabulary. In addition, we added special tokens to the start and to the end of each sentence, padded and truncated all sentences to a single constant length, and explicitly specified what are those padding tokens, using the ''attention mask''.\footnote{The attention mask is a binary tensor indicating the position of the padded indices so that the model does not attend to them.} We created an iterator for our dataset that helps saving memory during training and, by doing so, boost the training speed; then, we produced the BERT classifier and created an \emph{Adaptive Moment Estimation} (ADAM) optimizer, an efficient optimization algorithm based on gradient descent, with batch size $32$, learning rate $5e^{-5}$, and $2$ epochs.

\section{The Sampling Module}

% The sampling module takes as input the output of the estimation module, which are a list of numbers between zero and one that represent the probability of each proposal passing if there was a regular vote. For each input, the module outputs a sample-fraction, which is also a number between zero and one. Such outputs indicate the attention size for a particular proposal, i.e. the percentage of the population that should vote.

The goal of the sampling module is to maximize the quality of the decision made on each proposal while minimizing the amount of needed community attention. 
% Intuition, the larger the attention size, the greater the quality of decision making. Moreover, if the attention size is 1, meaning all members of the community will participate in the vote, we will make the same decision as if we held a regular vote, since we actually held one. Thus, in this case, the decision's quality would be the higher. 
In our design of the sampling module we aim at explicitly considering this trade-off between the attention used from the community and the quality of the decision making.

% \subsection{Our Sampling Algorithm}

% Increasing attention leads to improved decision quality. It is, however, important to keep in mind that the greater the attention, the greater the resources involved, and we want to minimize this. Thus, our goal will be to reach a compromise between the size of the attention and the quality of the decision, and then build (based on our goal) a mathematical function that returns the desired attention size for each probability obtained previously.

\begin{figure}[t]
    \centering
    \includegraphics[width=8cm]{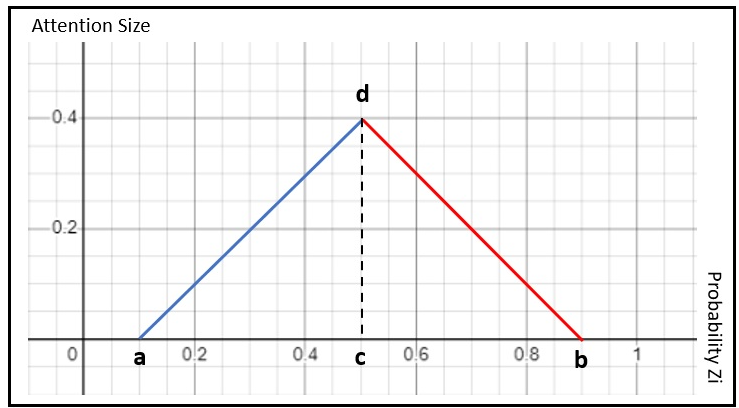}
    \caption{A triangle function as a sampling module: the $x$-axis represents the estimation of the probability of accepting a proposal if a regular vote were to be held (obtained from the estimation module), while the $y$-axis represents the sample size to be taken from the community for the proposal. %
    For the function shown here, if a proposal receives an estimated acceptance probability that is less than $a$ or greater than $b$, then the proposal is automatically rejected or accepted (respectively), and, in particular, no attention is needed from the community (as the value of the function at these regions is $0$); intuitively, this is so as, in such cases, the proposal is estimated to be almost in consensus. Between those points (i.e., for values between $a$ and $b$), the less estimated agreement there is on a proposal the larger sample size is selected to decide on it; hence the triangular shape.} 
    \label{figure:Triangle Function}
\end{figure}

Intuitively, the more in consensus a proposal is -- according to the estimation module -- the smaller the sample size can be; while, correspondingly, the less in consensus a proposal is -- according to the estimation module -- the larger the sample size can be:
  e.g., if the estimation module estimates the success probability of a proposal by almost $100\%$, then it is unlikely -- given that the estimation module is realized reasonably well -- that even a small sample size will be incorrect on the proposal; while if the estimation module estimates the success probability of a proposal to be, say, $51\%$, then even a high quality realized estimation module may be wrong, and, crucially, a small vote sample may also be wrong.
  
Corresponding to the intuition explained above, we consider a family of functions -- \emph{triangle functions} -- as shown in Figure~\ref{figure:Triangle Function}, that take as input a probability and outputs a sample size (as a fraction).

%Figure 2 illustrates the triangle's purpose: in this example, according to community's rule, if a proposal receives a probability of passing that is less than point a, the proposal is automatically rejected, so no attention needed from the community. In addition, the community stipulates that if a proposal gets a probability to pass greater than point b, the proposal is automatically accepted and therefore no attention is required in this case. As probabilities rise up to a certain limit - point c, attention sizes increase. If a proposal received a probability pass that close to point a, then it is most likely a proposal that is not "strong", so the logic says that not many people voted for it, and therefore a small sample is sufficient. In contrast, if a proposal receives a probability of passing that close to point c, it is likely that opinions about it are extremely divided and a relatively large sample is needed. The same idea continues when you cross point c. Probabilities closer to point c represent controversial proposals and therefore a larger sample must be used to decide, whereas Probabilities closer to point b represents a more powerful proposals and thus a smaller sample will be enough.

To completely realize the sampling module it is not enough to select the family of such triangle functions, but a specific triangle function has to be selected (consider Figure~\ref{figure:Triangle Function}).
According to the above intuition, as a larger sample size leads to better decision quality -- but also uses more of the community attention -- it would be unfair to compare different triangle functions that use different amounts of community attention, as the one that allows for comparatively larger sample sizes would outperform the other, less attention-demanding one.
Thus, to allow for an upright comparison of functions
it is first necessary to divide the family of triangular functions into subfamilies, one subfamily for each expected use of community attention. 

In our experiments (to be described in details in Section~\ref{section:experimental design}), we have several values of required average community attention, specifically: $0.1$, $0.2$, $0.3$, $0.4$, $0.5$; each such value in effect defines a subfamily of triangle functions that, given specific data -- and hence, given specific output values from the estimation module -- use this value-worth of average community attention.
Equivalently, to determine the relevant functions for a specific, given required average community attention (i.e., average fractional sample size), i.e., specific values of the points $a$, $b$, $c$, and $d$, we proceed as follows.
First, for simplicity, we refer only to the case where $a = 0$ and $b = 1$, i.e., there is no probability value $z_i$ that below it a proposal is to be automatically rejected or above it a proposal is to be automatically accepted (equivalently, with a sample fraction of $0$). This means that a specific triangle function can be represented by a tuple of $(c, d)$.
Second, for a given average community attention value $q$ from the ones mentioned above, we repeat the following process $100$ times, to get $100$ different triangle functions whose average community attention equals to $q$:
  we sample a value for $c$ uniformly at random between $0$ and $1$ (i.e., $c \sim U(0, 1)$); then, we do a binary search to find a value for $0 \leq d \leq 1$ such that the average community attention of the triangle function represented by $(c, d)$ corresponds to $q$, with respect to the distribution of the outputs of the estimation module (for a specific test data, as described in Section~\ref{section:experimental design}). As a result, we have $100$ triangle functions whose expected average community attention equals $q$, for each $q \in \{0.1, 0.2, 0.3, 0.4, 0.5\}$.

\section{Experimental Design}\label{section:experimental design}

We describe the datasets used for the computer-based simulations as well as the specific computations and evaluation metrics that were used.

\subsection{Datasets}

In order to examine our architecture, we used real data collected both from Kaggle, a popular machine learning site, and Snapshot, a popular DAO voting platform.

\paragraph{Kaggle data}
Kaggle is a platform that hosts data science competitions for business problems, recruitment, and academic research purposes~\cite{bojer2021kaggle}. As part of that, Kaggle find and publish datasets.

Kaggle was used to find similar data that could be treated as proposals and decision-making. Our goal was to find a dataset contains text-type data that we could use to run the NLP methods on. Moreover, we searched for data that has binary tags. 

The data we have chosen is the \emph{Research Articles} dataset.\footnote{\url{https://www.kaggle.com/datasets/vetrirah/janatahack-independence-day-2020-ml-hackathon}}
It consists of an abstract and a title for a set of research articles and the purpose is to use it to predict the topic of each article. The research articles are sourced from the following 6 topics: Computer Science, Physics, Mathematics, Statistics, Quantitative Biology and Quantitative Finance.

In order to use the above dataset, we had to choose a topic that would serve as a label; furthermore, ideally, that topic should be relatively balanced. We chose the topic ``Computer Science'' to represent our label:
  the label is $1$ if the article is associated with this topic and $0$ otherwise.

There are $20,972$ samples in the dataset, but due to computing power constraints, our dataset contained only $1000$ samples. The training set contained $85\%$ of the dataset (i.e. $850$ samples) while the test set comprised of $15\%$ (i.e. $150$ samples). Almost $41\%$ of the training set and $46\%$ of the test set has the tag $1$; i.e., the data is relatively balanced. To use the data for our needs, we consider $1$ as an expected approval of a proposal and $0$ as its rejection.

\paragraph{Snapshot data}
Snapshot is a popular tool used by certain Decentralized Autonomous Organizations (DAOs) that allows for community using tokens. 
The platform allows for multiple voting systems - Single choice, Approval voting, Quadratic voting, and more. Considering the platform is an open source project, it provides access to information of participating digital communities, including published proposals and votes.

To create a relatively large data set, we combined data from four organizations: \textit{Balancer} -- an automated portfolio manager and trading platform; \textit{YAM Finance} -- engaged in DeFi projects; \textit{Aavegotchi} -- a DeFi-enabled crypto collectibles game; \textit{Aave} -- a decentralized finance protocol.
Filtering has been done so that only binary proposals, with the option of accepting or rejecting them, were considered. 

There are $499$ samples (after the filtering process) in the dataset. The training set contains $85\%$ of the dataset (i.e., $424$ samples) while the test set comprises of $15\%$ (i.e., $75$ samples). Almost $86\%$ of the training set and $83\%$ of the test set is with tag $1$; i.e., the data is relatively unbalanced. (We will address this issue later by selecting certain metrics for testing the algorithms.)

\subsection{Evaluation Metrics}

We describe the evaluation metrics used for the estimation module, for the sampling module, and for the overall realizations of the architecture.

\paragraph{The Estimation Module}
In the estimation module, each classifier produced a set of probabilistic estimations for whether each of the proposals would have accepted. For evaluating and comparing the different algorithmic realizations of the estimation module, we have chosen the following metrics:
\begin{itemize}

\item
\textbf{accuracy}:
This metric is a classic metric that is equivalent to the proportion of the number of signals that were predicted correctly to the total of the input samples~\cite{hammad2021myocardial}. This metric provides an overoptimistic estimate of the classifier ability on the majority class and therefore suitable only to balance data~\cite{chicco2020advantages}.

\item
\textbf{The F1 score}:
This metric consists of the \emph{precision}, which is the proportion of the true positive samples of the overall predicted positive observations; and \emph{recall}, which is the portion of the true positive samples from the overall predicted negative observations. F1 score is especially useful for imbalanced data~\cite{hammad2021myocardial}, and is formally as follows:
$\textrm{F1-score} = \frac{2 \cdot (Recall \cdot Precision)}{Recall + Precision}$.

\end{itemize}

To evaluate our models we used the accuracy for the Kaggle dataset -- which is balanced -- and the F1-score for the Snapshot dataset -- which is imbalanced.

\paragraph{The Functions Quality}
As described earlier, we evaluated the sampling module for five subfamilies of triangle functions, with a size of $100$ each, where each subfamily corresponds to one of the five values of the average community attention $q \in \{0.1, 0.2, 0.3, 0.4, 0.5\}$. 
To determine which function is best for each subfamily, we define the quality of a solution, as follow. 

Consider $n$ proposals and a list of estimated
probabilities (obtained from the estimation module), $z_j$, $j \in [n]$. 
Consider a triangle function $f$ whose outputs on those $z_j$ values is $t_j$, $j \in [n]$ (so $t_j = f(z_j)$, $j \in [n]$). 
Recall that the decisions $\widehat{y}_j$, $j \in [n]$ will eventually be made by majority vote of a randomly selected subset of $t_j$, for proposal $j$. 
Consider the correct decisions $y_j$, $j \in [n]$, which correspond to a full vote on proposal $j$.
Then, the quality of a solution (represented by $\widehat{y}_j$, $j \in [n]$ is defined to be the fraction of computed decisions that are consistent with the correct decisions; formally:
\[I_i= \begin{cases}
    1, \textrm{if } \widehat{y}_i=y_i \\
    0, \textrm{otherwise}
        \end{cases}
\]
\[quality = \frac{\sum_{i=1}^{n} I_i}{{n}}\]
Due to the random selection of votes (based on the sample size), it is important to evaluate the average quality. Functions with the highest value are the best for each subfamily.

\section{Results}

We describe and discuss our results for the estimation module, for the sampling module, and for the architecture as a whole.

\subsection{The Estimation Module}

\begin{figure}[t]
    \centering
    \includegraphics[width=8cm]{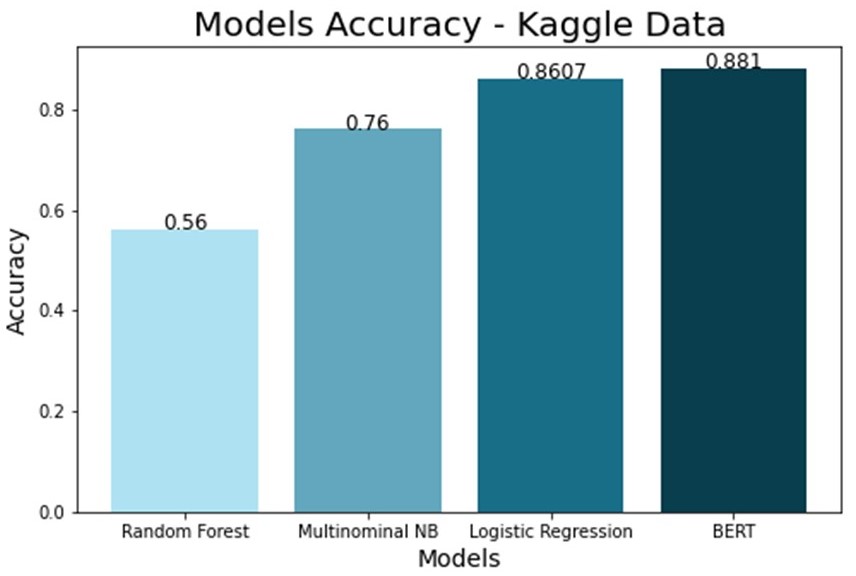}
    \caption{Performance For the Kaggle Data Set according to the Accuracy metric: the $x$-axis represents the four models we have used - the three on the left are classic classifiers and the fourth is a language model. In the $y$-axis, the accuracy index values for each model are presented.}
    \label{figure:Performance ACC}
\end{figure}

\begin{figure}[t]
    \centering
    \includegraphics[width=8cm]{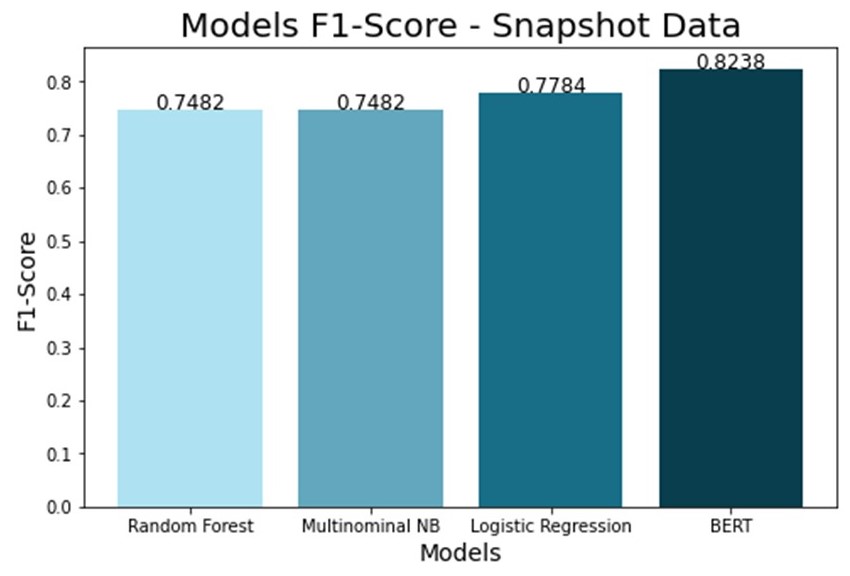}
    \caption{Models Performance For Snapshot Data Set According to the F1 Score  Metric: the x-axis represents the four models we used - the three on the left are classic classifiers and the fourth is a language model. The y-axis represent the F1 score index values for each model.}
    \label{figure:Performance F1}
\end{figure}

ML algorithms, both classical and language model, were used in the first stage of the architecture. Based on the data set characteristics, the way to measure those algorithms performance was chosen. 

On the Kaggle data set the algorithms were examined according to the accuracy index.
Among all the models tested, the language model (BERT) has the highest index value as shown in Figure~\ref{figure:Performance ACC} but not by a large margin from the Logistic Regression model, which leads all the other classical models. The results are consistent with the fact that a language model encompasses both words and context, so it is more developed and may lead to better predictions. 

However, on the Snapshot data set the algorithms were examined according to the F1 score index. It is evident in Figure~\ref{figure:Performance F1} that there are no huge differences in the quality of predictions between the three classical algorithms. Random Forest and Multinominal NB received identical index values while the index value of the Logistic Regression was slightly better. In this study as well, BERT has received the highest index value. However, BERT's results are not very high. As mentioned earlier, the data source is a platform for community members to formulate their own proposals, so the data set may contain low-quality text which can reduce the metric value.

\subsection{The Sampling Module}

\begin{figure}[t]
    \centering
    \includegraphics[width=9cm]{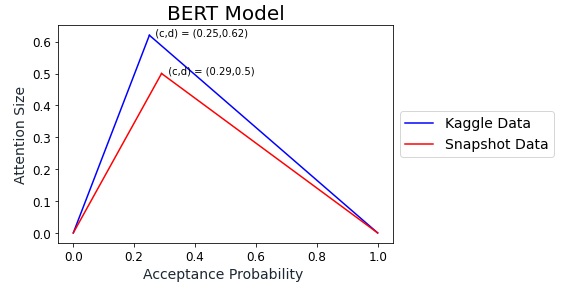}
    \caption{Example of the chosen triangle function for bERT with fixed average attention size of 0.4: the x-axis represents the estimation of the probability of proposal acceptance and the y-axis is the attention size needed for the vote. Here is an example of the BERT triangle function for a fixed average attention size of 0.4 - the blue triangle is for the Kaggle data set and the red triangle is for the Snapshot data set.}
    \label{figure:Functions}
\end{figure}

\begin{figure}[t]
    \centering
    \includegraphics[width=12cm]{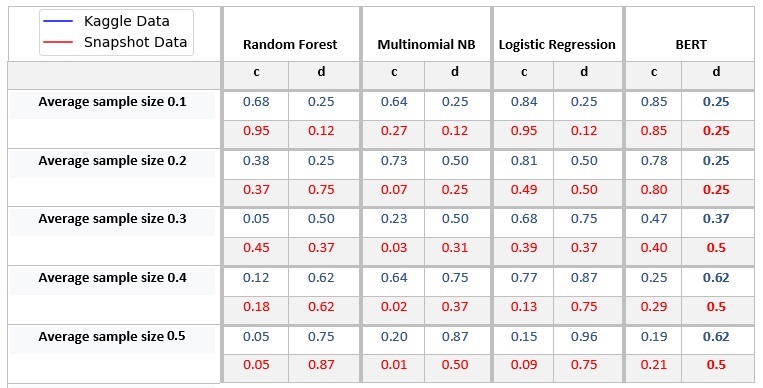}
    \caption{The functions selected for each criterion value and for each algorithm: the left column in the table represents the criterion value - average sample size. For each algorithm the values (c, d) of the selected function are represented. The blue values refer to the Kaggle data set and the red to the Snapshot data set.}
    \label{figure:c and d}
\end{figure}

In the Sampling Module stage, the functions selected based on the probabilities from the previous phase (ML module). Using five fixed attention sizes (0.1, 0.2, 0.3, 0.4, 0.5), we selected the functions that were most suitable under the conditions we assumed. Since the functions are selected by the probabilities from the ML module stage, there are such functions for each classifier. Figure~\ref{figure:Functions} illustrates an example of these functions. It seems from Figure~\ref{figure:c and d}, which represents the (c,d) values of the best functions, that for both data sets, in most cases, the higher the average sample size, the higher the d value. C value, on the other hand, does not show any trend.

\begin{figure}[t]
    \centering
    \includegraphics[width=8cm]{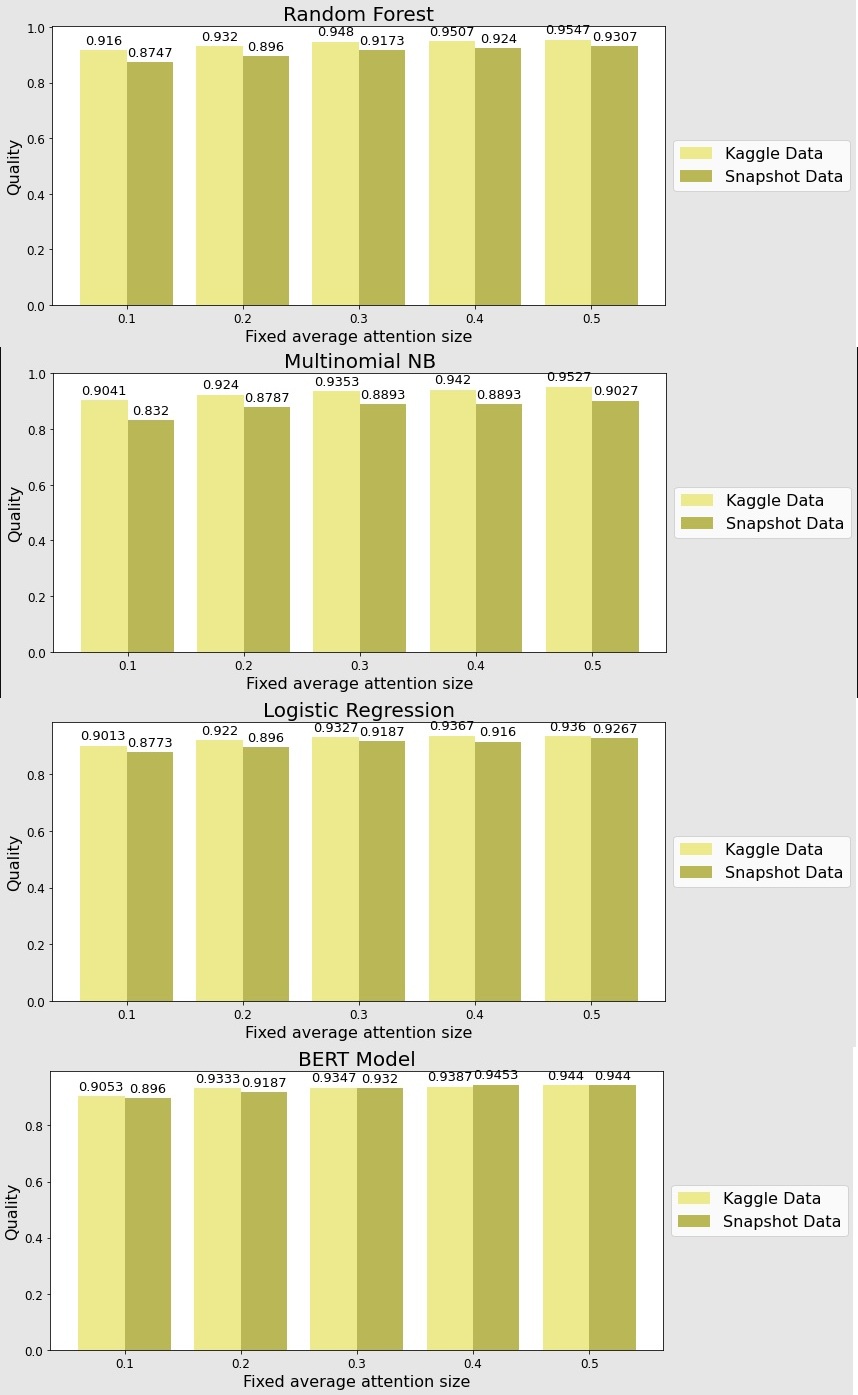}
    \caption{Quality VS Fixed Average Attention Size:  the x-axis represents fixed average attention size that was determined. The y-axis is the highest quality (how accurate was the model) that the model received.}
    \label{figure:Quality}
\end{figure}

\subsection{Quality Of The Overall Architecture}

For each classifier from the ML module phase and for each fixed average attention size, the appropriate function for the sampling module phase was found. This function was chosen because it gave the highest quality among the functions tested. Based on Figure~\ref{figure:Quality}, we can see that the quality ratio, that is, the proportion between the amount of decisions that were consistent with the original decisions and the total decisions, is relatively high. Moreover, when examining quality against the fixed average attention size, there is an upward trend.

\section{Conclusions, Limitations, and Outlook}

Attention-aware social choice refers to tackling the conflict rooted at the desire of a community to include its participants in the decision making processes despite their limited time and attention. 
Our work examines a possible solution that eventually can be used by digital sovereign communities that face this problem; it proceeds by a combination of two techniques: Natural Language Processing (NLP) and sampling. The system initially include a trained NLP model, to which the governance proposal to change the status quo are sent, that predict the probability that the proposal would pass if a regular vote were held. Based on these probabilities, the sampling module determines the number of participants that each proposal will receive, whose votes will be tallied and a majority decision will be made based on these votes.
We have shown the feasibility of our architecture by implementing several realization of the proposed architecture and evaluating them using computer-based simulations on real-world data. 

We do have some limitations, though.
Next, we discuss some conclusions from our work (Section~\ref{section:conclusions}); some of its limitations (Section~\ref{section:limitations}); and, finally, some avenues for future research (Section~\ref{section:future}).

\subsection{Conclusions}\label{section:conclusions}

We discuss some conclusions stemming from our work:
\begin{itemize}

\item
Given the relatively similar quality of using the simple NLP model and the more sophisticated one, it seems that, in some sense, the more sophisticated one is an unnecessary complication for some settings.

\end{itemize}

\subsection{Limitations}\label{section:limitations}

We discuss some of the limitations of our work:
\begin{itemize}

\item
Our method may suffer from bias. In particular, as we use learning algorithms that learn from past decisions towards the prediction of future decisions, we may use a learning algorithm that is skewed and biased towards the errors of the past.

One possible remedy to this problem may be societal:
  Whenever there is a very controversial decision,
  the community may inform the algorithm to down-bias its learning.
A complete solution along these lines may naturally be more involved, though.

\item
Our method may open new attack vectors on the governance system. In particular, a malicious agent may submit their proposal over and over, hoping for the randomness in the prediction algorithm to come in her favor once in a while.

While a careful analysis is in place, one immediate -- albeit only partial -- solution may be to use the prediction algorithm only as a suggesting tool and let a different institution (e.g., a pre-selected agent committee) make the sampling decisions.

\end{itemize}

\subsection{Outlook}\label{section:future}

Below we discuss some avenues for future research.
%
% Due to the sensitivity of the occupation and the risk that a proposal could pass when it was not supposed to, it is imperative to undertake further work that maximizes the capabilities of the system and minimizes the system's risks.
%
\begin{itemize}

\item
\textbf{Further comparisons}
We have concentrated on one family of (triangle) functions. Naturally, more families may prove to be more suitable for different settings; e.g., more continuously-looking Gaussian-shaped functions.

\item
\textbf{Further realizations of the architecture}:
Here we proposed several realizations for the different modules that comprise the overall architecture. Developing further realizations for the estimation module, the sampling module, and the decision module is a natural future direction of research.
This include, in particular, more conservative solutions:
  Note that, here, we defined the quality of a solution as simply the fraction of proposals on which the solution is correct (with respect to the correct decisions). In certain cases it may be more desirable to optimize a more ``conservative'' solution, as a proposal that is accepted even though it should not have been may be more damaging than vice versa. One possible solution for this would be to evaluate a different realization of the decision module, e.g., one that requires a supermajority for a proposal to be accepted.

\item
\textbf{A dynamic architecture}:
We envision a system based on the architecture presented here to be used by communities in a dynamic way. I.e., while the treatment here considers a static set of proposals -- and thus the required average attention can be a-priori known -- in the real world a dynamic algorithm to set the specific triangle functions of the sampling module to be used for each incoming proposal shall be dynamically be computed.
Related, a dynamic generalization of the architecture may use a self-improving realization, e.g., by using reinforcement learning to tune the realizations of the different modules as more proposals arrive. (A more advanced possibility is to pick, once in a while, artificially large sample vote sizes to re-train the NLP models.)

\item
\textbf{Incentivizing participation}:
If and when a tool based on the architecture outlined here is to be employed in the real world, certain issues can arise.
First, here we implicitly assumed that the community is very ``disciplined'', in the sense that when the sampling module declares a required sample size indeed there are sample-size-many voters who show up to vote. Practically, some economic incentivization may be needed.

\end{itemize}

\bibliography{bib}
\bibliographystyle{plain}

\end{document}